# Publish *and* Perish: Creative Destruction and Macroeconomic Theory[1]

Jean-Bernard Chatelain[2] and Kirsten Ralf[3]

**Abstract**

A number of macroeconomic theories, very popular in the 1980s, seem to have completely disappeared and been replaced by the dynamic stochastic general equilibrium (DSGE) approach. We will argue that this replacement is due to a tacit agreement on a number of assumptions, previously seen as mutually exclusive, and not due to a settlement by 'nature'. As opposed to econometrics and microeconomics and despite massive progress in the access to data and the use of statistical software, macroeconomic theory appears not to be a cumulative science so far. Observational equivalence of different models and the problem of identification of parameters of the models persist as will be highlighted by examining two examples: one in growth theory and a second in testing inflation persistence.

JEL classification numbers: B22, B23, B41, C52, E31, O41, O47.

Keywords: Macroeconomic theory, controversies, identification, economic growth, convergence, inflation persistence.

« *'Nothing is vanity; on to science, and forward!' cries the modern Ecclesiastes, that is to say Everyone in the world. And yet, the cadavers of the bad and the lazy fall on the hearts of others... Ah! quick, quick a bit, over there; beyond the night, these future, eternal, rewards... do we escape them? ...* » Rimbaud A. 1873, A Season in Hell, Lightning.

## 1. INTRODUCTION

Mainstream macroeconomic theories of the 80s, such as monetarism, real business cycle theory, fixed-price disequilibrium theory, and *ad hoc* dynamic models not based on optimal saving, seem to have completely disappeared and belong now to the history of economic ideas. They were replaced by the dynamic stochastic general equilibrium (DSGE) theory. Nonetheless, controversies on macroeconomic theory were not settled and scientific progress is not achieved: «For more than three decades, macroeconomics has gone backward» (Romer 2016, 1). This paper investigates what are possible reasons for the adoption of the DSGE model and why it will probably not be the end of disagreement in macroeconomics, especially with respect to the gap between normative and positive economics.

---

[1] We thank Michel de Vroey and two anonymous referees for very helpful comments.
[2] Paris School of Economics, University Paris I Pantheon Sorbonne, PSE, 48 boulevard Jourdan, 75014 Paris. Email: jean-bernard.chatelain@univ-paris1.fr
[3] ESCE International Business School, 10, rue Sextius Michel, 75015 Paris, Email: Kirsten.Ralf@esce.fr.

Latour (1988) puts forward that controversies are a typical step of 'science in the making'. Long-lasting scientific controversies, however, signal a failure to accumulate knowledge. Macroeconomic theory appears to be a case where data seem to be unable to settle controversies, such as adaptive expectations *versus* rational expectations (Pesaran 1981, 1987). Other factors than data may decide on what could be the prevalent theory at a given point in time (Colander 2009). Hendry (2005) emphasizes that theory driven approaches, since their credibility depends on the transient credibility of rapidly evolving theory, induce transient and non-structural evidence, «But postulating an endless sequence of models and theories that get rejected in turn fails to incorporate learning from the evidence.»

Since one of the aims of macroeconomics is to give policy advice, macroeconomics is prone to controversy because of conflicting interests in society. The political opinion of the researcher or the belief in market forces may influence the way he derives his economic model. Furthermore, macro-econometrics faces more difficulties than micro-econometrics, not only because macro-econometrics looks at several markets instead of a single one, but also due to the difficulties that arise from the observational equivalence of different theories and from parameter identification and endogeneity issues. Identification issues are at the juncture between two broadly defined distinct methodological approaches of macroeconomic theory: normative, theory-first or theory-driven (Hendry 2005, Juselius and Franchi 2007, Spanos 2009) and Walrasian approach (De Vroey 2016), such as New-Keynesian DSGE models, *versus* positive, data-first or Marshallian approach, such as structural vector auto-regressive (SVAR) models (Kilian and Lütkepohl 2017). These identification issues have been highlighted among others in Pesaran (1981), Beyer and Farmer (2004, 2008), Canova and Sala (2009), Blanchard (2016), and Romer (2016). Marchionatti and Sella (2017) argue that it is the Neo-Walrasian legacy of DSGE models that led to the poor predictive performance and the interpretational weakness of these models.

We highlight a number of facts why DGSE models avoid falsification, so that data cannot settle controversies. Pesaran (1981, 1987), Binder and Pesaran (1995), Canova (2009 and 2012), and Koop *et al.* (2013) demonstrate that the solutions of rational expectations 'New Keynesian' DSGE models are observationally equivalent to adaptive expectations 'Old Keynesian' vector auto-regressive (VAR) models. It is not possible to test one theory against the other one when their predictions are identical, although they use different hypothesis. Hence*, it is impossible to settle the controversy of rational expectations against adaptive expectations using data.*

It is then usually argued that DSGE models are superior to adaptive expectations VAR models because they are based on intertemporal microeconomic foundation. The theoretical discipline of micro-foundations is empirically justified by the endogeneity of macroeconomic policy instruments leading to parameter identification problems (Lucas 1976).[1] Omitting the endogeneity of the policy instruments may lead to biased parameters of macroeconomic policy transmission equations. This endogeneity bias may change if the policy-rule parameters determining the endogeneity of the policy instruments change. This specification error of macroeconomic policy transmission mechanisms may lead to policy mistakes. Kilian and Lütkepohl (2017), section 6.3.6, argue that reduced form structural VAR may also conduct policy analysis taking into account the endogeneity of policy rules. But micro foundation does not solve *all the remaining* identification and observational equivalence issues when testing macroeconomic theory.

---

[1] It has been argued, however, that the assumption of a single representative agent does not represent a true microeconomic foundation since all heterogeneity of agents is excluded. But this discussion is beyond the scope of the present article, see e.g. Marchionatti and Sella 2017 and Vercelli 1991.

Some micro-foundations may imply *too many structural parameters* with respect to reduced form parameters, i.e. *under-identification* of the statistical model used to test the theory (Canova 2012). When facing a problem of parameter identification, a useful normative macroeconomic theory can be a useless positive macroeconomic theory. In what follows we will present the example of empirical studies of convergence where optimal growth was not - and could not be tested against the Solow growth model, because the key parameter of preferences of optimal growth was not identified.

Furthermore, some micro-foundations correspond to *misspecified* models. For example, they do not fit the hump-shaped impulse responses describing the *persistence* of macroeconomic time-series obtained with adaptive expectations vector auto-regressive models including two lags (Estrella and Fuhrer 2002). In order to fit persistence, new-Keynesian DSGE models are hybrid models mixing rational expectations and *ad hoc* adaptive expectations. They add *ad hoc* lagged dependent variables (inflation and habit persistence, for example) and/or *ad hoc* auto-correlation parameters of shocks, which are not grounded by micro-foundations. As an example, we compare these two observationally equivalent models with an observationally equivalent VAR model with adaptive expectations. Because the persistence parameters are *ad hoc* in the New-Keynesian model and not derived from micro-foundations (criterion 1), we propose three alternative criteria other than micro-foundation to compare the three models of persistence. The VAR model with adaptive expectations includes fewer parameters (criterion 2), some of the parameters of the New-Keynesian DSGE model may be not identified (criterion 3) and the interpretation of some parameters is opaque, complicated, and counter-intuitive (criterion 4).

Additionally, micro-foundations add *opacity on identification issues* because reduced form parameters are highly non-linear functions of preferences and technology parameters. «*The treatment of identification now is no more credible than in the early 1970s but escapes challenge because it is so much more opaque.*» (Romer 2016, 1) These technicalities have been handled in Canova and Sala (2009), Iskrev (2010), Komunjer and Ng (2011), Canova (2012), Koop *et al.* (2013). So far, they had no feedback on *changing* the specifications of New-Keynesian DSGE theory in order to reduce weak identification issues.

Successful theories are designed to many allies among researchers. A successful normative, theory-first macroeconomic theory is likely to satisfy this necessary Darwinian survival of the fittest criterion, in order to be protected from its empirical falsification: *Design 'not even wrong' observationally equivalent theories to pre-existing theories, which may include too many parameters for identifying them with available data and which may face opaque weak identification issues.* Hence, controversies on parameters *cannot* be settled. These theories allow multiple interpretations, numerous variations and combinations in subsequent papers, controversies, citations and fame. These theories are likely to be relatively more sophisticated and to attract many allies among bright scholars. Ockham's razor of parsimony of the number of parameters of a theory is not rewarding for academic careers. It removes too many opportunities of publications and citations.

There is a complementarity between the persistence of controversy in macroeconomics and macro-econometrics and the researcher's incentives to preserve the theoretical 'discipline' of their theory-first approach, while *deliberately selecting observationally equivalent theories and forgetting discipline with respect to identification issues when using econometrics.*

Our plan is as follows: In the next section we argue that four competing theories of the 1980s ran out of fashion and have been replaced by dynamic stochastic general equilibrium models. In section 3, we distinguish a normative and a positive approach to economic questions with optimal growth theory as an example. In a fourth section, we compare the adaptive expectations vector auto-regressive model and the rational expectations DSGE model with respect to the problems of observational equivalence and identification of parameters. In section 5, we develop some criteria to decide which of different competing theories to take and try to make a forecast for future of the DSGE model. Section 6 concludes.

2. EX-FAN OF THE 1980S: THE ACCELERATED DEPRECIATION OF MACROECONOMIC THEORIES

*2.1. Dead macroeconomic theories*

Leijonhufvud (2006) and De Vroey (2016) use the movement of climbing a decision tree as a metaphor to describe the change of thought in macroeconomics. Each fork is seen as a decision to pursue the one or the other idea. Sometimes researchers get stuck on one of the lower branches. Then, they backtrack to an earlier bifurcation, restart another approach, and the dead-end theory dies. A number of indicators, besides the judgments of researchers themselves and historians of macroeconomics, may measure the decreasing interest of theories: They are not used or cited in current research (see appendix), no recent PhD thesis is built on them, they are not used for current policy advice, they are neither mentioned in current master level textbooks, nor in current undergraduate level textbooks, they are mentioned mostly by historians of economic thought. This loss of interest may be sudden or gradually, a few ones may even be resurrected.

We will briefly introduce four deceased mainstream macroeconomic theories of the 80s with an extract from their respective obituaries. In doubt, ask yourself: If a master student comes to your office and argues that he would like to do a PhD using one of the following 80s theories, would you accept to be his PhD advisor?

2.1.1. Monetarism

Monetarism stressed the importance of policy rules for monetary aggregates in order to fight inflation, based on the quantitative theory of money, an old economic idea with already an impressive literature. Emerging in the 50s, it was American-based with a prominent Chicago Nobel prize winner with political clout, institutional support in the Fed, the Bundesbank, and other central banks. Monetarism is credited for the monetary policy success of the sharp disinflation of the early 80s in the US, the UK, and continental Europe. Since the 2000s, no PhD in monetary economics hired in a Central Bank research department, refers to Friedman's monetarism. The official obituary notice is short: *The estimate of the velocity of money is not stable* (see Taylor's (2001) interview with Milton Friedman as stated in De Vroey (2016)). A Taylor rule on the federal funds rate replaced Friedman's monetary rules. In the 80s, monetarists would have ironically grinned at forecasts (nobody made) of the death of such a powerful theory in the 90s. A resurrection of monetarism in the near future is unlikely. Quantitative easing seemingly led to negligible increases of inflation and of real output in Japan, US, UK, and the Euro-area in recent years.

2.1.2. *Ad hoc* rational expectations or Keynesian models not based on optimal savings

Regardless of their school of thought and the varieties of their ideas, a large market share of macroeconomic models in the 80s were static or dynamic models not based on optimal savings derived from intertemporal microfoundations (for example, Dornbusch's (1976) rational expectations model of the overshooting of exchange rate). It has been somehow forgotten that, not only Keynesian models, including large scale forecasting models in central banks and treasuries, but also many small scale rational expectations models were not based on intertemporal optimal savings at the time. In the 1990s, optimal savings based on the intertemporal substitution effect of the interest rate on consumption, with infinite (Ramsey) or finite horizon in overlapping generation models, turned out to be of compulsory use in mainstream macroeconomics. This was fostered by theories of endogenous growth, of the open economy, and of real business cycles. The obituary notice says: *Either they were static models with optimal choice or they we dynamic models lacking intertemporal microeconomic foundations. Any macroeconomic idea which cannot be introduced into a dynamic model assuming Ramsey optimal saving infinite horizon (overlapping generation model) or in infinite horizon for a proportion of households is a dead idea.*

2.1.3. Disequilibrium macroeconomics assuming quantity rationing on goods and labour market

As a subset of the models of section 2.1.2, this theory assumed price rigidity with excess supply or excess demand using static models and econometrics. It was French and Belgium-based without a Nobel prize awarded. The obituary notice mentions: *It did not find allies among US-based macroeconomists.* Researchers and their PhDs shifted to endogenous growth theory or overlapping generation (OLG) models or macro-econometrics or real business cycles or microeconomics or retired.

2.1.4. Real business cycles

Kydland and Prescott's (1982) real business cycle (RBC) theory claimed monetary policy ineffectiveness while rejecting econometrics. It was referred to as 'dark ages' by John Taylor. It was American-based with two future Nobel-prize winners. It is possible that a bureaucratic labour market of policy makers has killed RBC theory. As the Fed and later the ECB and other central banks have a large market share of jobs for PhDs in monetary macroeconomics it was not sustainable in the medium run to hire staff and consultants only able to produce variations of RBC denying any use of central banks. Due to these circumstances, there is no obituary notice. However, Romer (2016) and De Vroey (2016) consider that there is a continuity between RBC and New-Keynesian DSGE models. This is further discussed in the next section.

Graduate-level macroeconomic theories which emerged in the 70s and flourished and prospered in the 80s faced depreciation in the 90s. A few economists, however, emphasized the surprising resilience of the 40s IS-LM model in undergraduate macroeconomic textbooks to understand economic policy during the 2007... economic crisis.

Where does this lack of cumulative knowledge come from? Are modern macroeconomic theories only driven by creative destructions, by new theories, doomed to be destructed later on? Are they, as asset price bubbles, only driven by fads and fashions for a given generation of macroeconomists with a time-period of 25 years?

*2.2. New-Keynesian DSGE theory*

Starting in the 1990s, New-Keynesian DSGE models drove out the above-mentioned theories. Is this a change of paradigm or just a transitory phenomenon? Drakopoulos and Karayiannis (2005), Duarte (2016) and De Vroey (2016) recommend to limit calling any creative destruction of paradigms a scientific revolution in the sense of Kuhn. The creative destruction of short-lived attempts to expand knowledge, with limited scope of explanation and with fragile empirical validation, is not equivalent to a major scientific revolution. The New-Keynesian DSGE theory is rather a mutation or a collage of existing theories. It resulted from a bargain between a few researchers in the real business cycle group and in the New-Keynesian group between 1992 to 1997, at the expense of Friedman's monetarism, of Prescott's monetary policy ineffectiveness, and models not based on Ramsey optimal savings. It is a combination of five existing elements:

(1) Kydland and Prescott's (1982) exogenous auto-regressive forcing variable, adding as many exogenous serially correlated forcing variables as time-series used in the estimation of New-Keynesian DSGE models.

(2) Calvo's (1983) staggered price-setting, leading to the New-Keynesian Phillips curve, where the expectation of future inflation is negatively correlated to an increase of the current output gap, as opposed to the accelerationist Phillips curve.

(3) Ramsey's (1928) optimal saving, where the future consumption is expected to increase following a rise of the real interest rate. The monetary policy transmission mechanism is based on this intertemporal substitution effect of the interest rate instead of the cost of capital.

(4) A Taylor (1993) rule: the interest rate set by monetary policy responds to the deviation of inflation from its target rate and to an output gap. Monetary aggregates are not necessary to explain monetary policy.

(5) Blanchard and Kahn's (1980) unique solution for dynamic systems, with the *ad hoc* assumption of the policy instrument (interest rate) and the policy targets (inflation and output gap) to be simultaneously forward-looking variables.

The core of the bargain was on agreeing to assume simultaneously assumptions (1) and (2):

On the real business cycle side, King agreed to include assumption (2) *ad hoc* price rigidity for a proportion of firms over time (Calvo's (1983) price setting), which the Real Business Cycles group had doctrinally refused to assume for the ten to fifteen years before.

On the New-Keynesian side, Rotemberg, Woodford, and Galí agreed to include assumption (1) *ad hoc* auto-regressive and possibly negative productivity shocks (Kydland and Prescott (1982)), which the New-Keynesian group had doctrinally refused to assume for the ten to fifteen years before.

While replacing the four theories mentioned above, New-Keynesian DSGE theory picked some of their assumptions and rejected others:

**Monetarism:** The New-Keynesian DSGE theory claims that monetary policy matters but not monetary aggregates. For the monetary policy transmission channel, it substituted a dynamic version of the quantitative theory of money (monetary aggregates have an effect on future output and prices) by the New-Keynesian Phillips curve. For the monetary policy rule, it substituted any type of monetary rule by the Taylor rule, using the interest rate as a policy instrument responding to the output gap and to the deviation of inflation from its target.

*Ad hoc* **macroeconomic models without optimal saving:** The New-Keynesian DSGE theory assumes that savings behaviour is derived from Ramsey intertemporal optimization. In downturns, the poorer the representative household is and the lower its relative fluctuation aversion, the larger are his savings in order to get back to its equilibrium stock of wealth. In booms, the richer the representative household is and the lower its relative fluctuation aversion, the lower are his savings in order to get back to its equilibrium stock of wealth.

**Disequilibrium macroeconomics:** The New-Keynesian DSGE theory does not assume that there is disequilibrium in quantities (neither excess demand nor excess supply) on the goods and the labour market, although it includes Calvo's (1983) assumption that a proportion of firms exogenously face price rigidities in each period.

**Real business cycles:** De Vroey (2016) and Romer (2016) consider that New Keynesian DSGE modelers are in the continuity of real business cycles models. Both approaches assume a Ramsey optimal saving model, as did the endogenous growth models. This is, for sure, a very constraining hypothesis for macroeconomic theory. However, there are several important changes which also suggest the death of the genuine project of real business cycles: monetary policy matters, some of the shocks are demand or monetary shocks, instead of only supply and productivity shocks. They use econometric (Bayesian) estimation of some of the parameters instead of calibrating all parameters. They assume staggered price-setting and price rigidity. They state the *ad hoc* hypothesis that policy instruments are forward-looking when policy targets are forward-looking. This implies positive-feedback policy rule parameters, which are not derived from first principles with an optimal policy makers behaviour as opposed to Ramsey optimal policy. They expand the number of auto-regressive shocks to be equal to the number of forward variables (instead of using only one measured time series of the productivity supply shocks).

The New-Keynesian model is a bargain on two elementary theoretical assumptions and three other core elements between leaders of opinions in theoretical macroeconomics around 1997. This new model had no formal and mathematical difficulty, adding together components already available in the economic literature. To the five compulsory elements are added a large number of variations (credit frictions, search on the labour market, open economies, ...), a patchwork of already existing elements that sometimes do not go together well, especially if the target is to explain the relationship between macroeconomic policy and macroeconomic variables and to advise policy makers.

One example of elements that do not go together well is the combination of (4) and (5) with the *ad hoc* assumption that the policy instrument and the policy target are simultaneously forward-looking variables. This leads to the unbelievable mechanism of a positive-feedback policy rule in order to stabilize inflation. «*In New-Keynesian models, higher inflation leads the Fed to set interest rates in a way that produces even higher future inflation. For only one value of inflation today will inflation fail to explode.* » (Cochrane 2011).

This *ad hoc* positive-feedback policy rule is the *opposite* of the negative-feedback mechanism of the two original papers. Taylor's rule (1993) is a negative-feedback rule leaning against inflation spirals assuming that inflation is backward-looking (Cochrane 2011). Blanchard and Kahn's (1980) condition for a unique solution (determinacy) of *ad hoc* linear rational expectations systems is an extension of the unique solution of the optimal-control linear-quadratic regulator (Vaughan 1970) for optimal negative-feedback rules. They rename 'costate' variables (policy instruments) of the Hamiltonian system in Vaughan (1970)

'forward-looking' variables of their *ad hoc* linear system, and 'state' variables (policy targets) of the Hamiltonian system 'backward-looking' variables. In the Blanchard/Kahn (1980) setting, it is up to the researcher to decide which variables are forward-looking and which are backward-looking.

Most macroeconomists abandoned their former 'school of thought' of the 80s. They were versatile and recycled themselves quickly to a new one. In despair of cumulative science, a few positivists may appeal to a nostalgia narrative: *There may remain a flavor of the spirit of 80s theory where we could decipher a far-fetched analogy, inheritance, influence in the New-Keynesian DSGE theory, with at least the reference to two common very broad stylized facts: price rigidity and monetary policy.*

### *2.3. Facts that do not go away?*

Duarte (2012) initially put forward Blanchard's view that stylized facts 'do not go away' as key driver of the New-Keynesian DSGE models' creative destruction of other theories. After detailed investigation, Duarte (2015) is more balanced. Stylized facts on business cycles remained controversial with trend *versus* cycle decomposition. Demand shocks *versus* supply shocks did not lead to clear-cut results. Duarte acknowledges that other factors than 'facts' may explain the New-Keynesian DSGE theory takeover.

Furthermore, Blanchard's arguments of 'facts that do not go away' as an explanation of the almost total takeover of DSGE models is challenged by a timing issue. The empirical evidence of monetary policy effectiveness and price stickiness was available at least in 1982, and not only in 1997. 'Facts that do not go away' was not sufficient to change the point of view of macroeconomists in the 80s. Concerning empirical observations a number of points are in order:

Monetary policy is effective. Statistics on inflation, federal funds rate, monetary aggregates, output, and unemployment during Volcker's disinflationary policy 1980-1982 were available and mentioned in the economic policy debates during these and the following years.

The stylized fact of staggered price rigidity was also known in the 70s and 80s. Go to a supermarket and check whether the prices of your favorite consumer goods change every week, even in the time of two-digit inflation.

The stylized fact that the econometric estimate of the velocity of money is not constant did not stop monetarism in the 70s and 80s.

The stylized fact that does not go away of housing price bubbles and of financial crisis (Reinhard and Rogoff 2010) was discarded in favor of the consensus of the New Keynesian DSGE synthesis before 2008.

The stylized fact that does not go away of an unstable relationship between inflation and output gap (Old-Keynesian *versus* New-Keynesian Phillips curve) was discarded in favor of the consensus of the New-Keynesian DSGE synthesis until now.

### 3. TESTING MACROECONOMIC THEORY

### *3.1. Normative versus Positive Theory*

In view of the change of macroeconomic theories and their ability to explain certain stylized facts, but not others, two approaches have been suggested in the literature. Either macroeconomic theory need not, cannot, and should not be tested or macroeconomic theory need, can, and should be tested.

In the first, normative approach, microeconomic, macroeconomic, and general equilibrium economic theories are only deductive theories, seeking e.g. the best allocation of scarce resources, as a branch of applied mathematics. *Deductive* theories need not, cannot, and should not be tested. For nearly two centuries, many economists, including Walras, Keynes, Debreu, and Austrian economists among others, did not design their theories in order to test them. Spanos (2009) describes this pre-eminence of theory approach. Economic theories are then only thought experiments that allow to derive those real-world phenomena that are considered to be most relevant by the researcher in a consistent and logical correct way from a number of assumptions. A theory is considered to be better than another one if it has fewer or less restrictive assumptions.

In the second, positive approach, economic theory is designed in a way that the parameters of the model can be estimated and their results can be tested. A rejection of the results should then lead to a rejection of the theory or at least to a major change of the economic modelling. This approach also requires that the parameters are identified. If two distinct sets of parameters of a theory predict exactly the same observations ('observational equivalence'), there is an identification problem for testing the theory. Additional data does not help to solve this problem, only a change of the underlying theory can lead to a uniquely identified set of parameters.

The above two views corresponded to a division of labour which prevailed until the 90s between macroeconomic theorists *versus* macro-econometricians, with each group being unable to master the distinct and highly specialized skills of the other group.

### 3.2. Controversies and tests of macroeconomic theory

Can observations of 'nature' settle the macroeconomic controversies, as in other fields of science? After Latour's (1988) 'science in the making' controversies, 'nature' may finally settle the controversy where knowledge accumulates into Latour's (1988) '*readymade science*' in the store of no-longer controversial '*scientific facts*'.

In Krueger's (2003, 189) interview, Malinvaud mentions: «*It was easy to study problems in microeconomic theory, which were well defined, and where a brain educated in mathematics could bring contributions. But macroeconomics was a more difficult domain to conduct research, because the questions were so involved and had many aspects. Let's say we were not on a clean land.*» Econometrics can be viewed as a branch of applied statistics. Microeconomics can be viewed as a branch of applied mathematics, dealing with economic choices based on applied optimization under constraints. Both of them are 'well-defined' and likely to follow a cumulative path for knowledge similar to the field of applied mathematics.

Macroeconomic theory, on the other hand, is related to many social factors: macroeconomic policy, political economy, the choice of the design of market institutions, ideology and ideas, the vested interests of social groups, structural changes brought by technical innovations, by history and by the geopolitics of power and conflicts against nations. Macroeconomics is prone, not only to country-period specific knowledge, but also to unstable and non-cumulative knowledge and potentially never-ending controversies.

Furthermore, it is more difficult to test macroeconomic theory than microeconomic theory, for a number of reasons:

(1) The correlations between macroeconomic time-series are not stable for periods longer than 10 to 20 years (40 to 80 quarterly observations). This is caused by structural changes such as breaks in the growth of productivity and technical change, changes of exchange rate regimes, changes of regulations on capital flows and cross-border banking and finance (financial globalization and its reversal), changes of political preferences for macroeconomic policies along with the learning process of structural changes in macroeconomic policy transmission mechanisms, changes of trade agreements and current account imbalances, demographic transition, wars, and increases of inequalities among others. Boyer (2015) describes theses combinations on the various markets as modes of regulation for given periods of time in each country, within which an accumulation regime occurs. These regimes may then be related to a monetary policy regime (Benati and Goodhart 2010).

(2) Small samples and small statistical populations. When not pooling time-series and cross-country observations, the number of observations remains low as opposed to large samples in microeconomic data. Structural change may reduce the relevant time dimension of time-series further. Natural experiments between comparable countries in order to evaluate causality of the effect of a policy treatment are frequent but the number of observations per country is very small. Statistical inference is therefore limited. One may compare North *versus* South Korea, Haiti *versus* the Dominican Republic, but it is hard to find more than two comparable countries.

(3) Natural experiments are not done in isolation: there are always several confounding effects. The determinants of economic growth may go up to fifty factors, with up to a dozen of different measures of each of the factors (up to 500 explanatory variables). A government budget cut appears at the same time as a GDP expansion, because exports increased in the short run due to a devaluation of the currency happening at the same time (for expansionary austerity in Nordic countries in the 80s). In Ireland in the 1990s, labour market reform occurs at the same time as foreign direct investment low-tax incentives and European Union subsidies.

(4) Macroeconomic time-series are persistent. They may have large auto-correlation parameters close to one ('unit-root'). This may lead to spurious regressions. Those spurious regressions can be published on the ground that the statistical power of unit root tests is small for small samples, such as forty years of quarterly data.

(5) All macroeconomic variables are endogenous (except geography and rainfall or information in the remote past). Testing macroeconomic theory faces the lack of identification, arbitrary identification restrictions, and weak identification using weak instrumental variables. Macroeconomic monetary and fiscal policy and their endogeneity and interactions with expectations complicate the estimations of macroeconomic policy effects.

*3.3. A useful normative theory can be a useless positive theory*

To highlight the problem of observational equivalence and identification we can take the example of a theory that predicts a linear relation between two variables $(x_t, y_t)$, where the number of reduced-form parameters, here $\beta$, is lower than the number of structural parameters. In the example below, the theory predicts that the weighted sum of two structural parameters denoted $(a, b)$ is equal to $\beta$:

$$y_t = \beta x_t + \varepsilon_{t+1}, \text{where } \beta = 2a + b \text{ and } \varepsilon_t \text{ is } i.i.d\ N(0, \sigma_\varepsilon^2).$$

From a *normative* theory point of view, this theory is *useful*, because one can explain comparative statics to undergraduate students. A 10% change of parameter $a$ *ceteris paribus* (with $b$ unchanged) has a different effect of $y$ on $x$ than a 10% change of $b$ with $a$ unchanged.

The theory may satisfy *useful* criteria and principles from a normative point of view, such as microeconomic foundations or rational expectations.

From a Koopmans or Cowles Commission positive economics point of view, this theory is *useless*, because it is impossible to distinguish using data the effect of $a$ (*ceteris paribus*) from the effect of $b$. It is also impossible to distinguish this theory from an infinite set of alternative theories that are *observationally-equivalent*, where the reduced-form parameter is any function of any finite or infinite number of structural parameters. This is such an obvious case of a useless under-identified model with too many structural parameters with respect to the number of reduced form parameters, that it is *never* mentioned in econometrics textbooks.

There is observational equivalence for the reduced-form parameter function of two structural parameters when *at least* two distinct sets of values of structural parameters $(a, b)$ provide the same prediction:

$$\hat{\beta} = 1 \Rightarrow a = \frac{\hat{\beta} - b}{2} = \frac{(1-b)}{2}.$$

For an estimated value of the reduced-form parameter $\hat{\beta} = 1$, there is an infinity of pairs of structural parameters $(a, b)$ such that $2a + b = \hat{\beta}$ that predict the same observations; take e.g. $b = 0$ and $a = 1/2$ or $b = 1$ and $a = 0$. To estimate one of the two structural parameters $(a, b)$, the researcher needs an additional theory justifying an additional identification restriction, such as $b = 2$. Then, he can estimate the other parameter: $\hat{\alpha} = (\hat{\beta} - 2)/2$. A *useful normative theory can be a useless positive theory with too many parameters*.

Since empirical observations alone are not enough to distinguish between different theories with the same estimated parameter values other selection criteria have to be found. One criterion has already been mentioned in Aristoteles physics twenty-five centuries ago, was taken up by Thomas of Aquino in 13th century, and is known in the literature as Ockham's razor, following William of Ockham (14th century): among all theories with exactly the same predictions, the simplest theory with the lowest number of parameters has to be chosen.[1]

The design of a positive theory has to go hand in hand with the precise design of the identification of its parameters in its empirical test, taking into account the availability of data. But academia may reward the quick and dirty 'imprecise' positive validation of a fascinating normative theory.

### 3.4. Testing optimal growth and convergence

As a first example of the under-identification of a normative theory which is observationally equivalent to a pre-existing theory, we will briefly discuss the estimations and tests found in the optimal growth literature. Cross-country convergence in the Ramsey-Cass Koopmans optimal-growth model in the 1990s was tested by regressing the growth of gross domestic product (GDP) per capita on an initial level of GDP/head. A meta-analysis found the estimated parameter on average around $\hat{\beta} = 2\%$ for the period 1960-2000 (Abreu *et al.* 2005). At least two competing theories may lead to the same reduced form regression.

---

[1] The divergence between normative *versus* positive knowledge facing Ockham's razor, being an old issue, is much broader than macroeconomic theory. For 'positive' science, God is an unnecessary hypothesis for the theory of celestial mechanics, according to Laplace's answer to Napoleon and for Darwin's theory of evolution. Aquinas' (1920, part 1, Q.2) second objection on the existence of God is: «Further, it is superfluous to suppose that what can be accounted for by a few principles has been produced by many. But it seems that everything we see in the world can be accounted for by other principles, supposing God did not exist.» But this does not imply that God does not exist. God is a necessary assumption for 'normative' or teleological theology (van Inwagen 2005, Glass 2016).

For the closed-economy model with a constant savings rate (Solow-Swan), the reduced form convergence parameter $\beta$ depends on four structural parameters: the elasticity of capital in the production function, $\alpha$, the growth rate of labour, $n$, the growth rate of labour-augmenting technical change, $x$, and the depreciation rate of capital, $\delta$ (Barro and Sala-I-Martin 2004, 112):

$$\beta(\alpha, x, n, \delta) = (1 - \alpha)(n + x + \delta) \text{ constant savings rate (Solow model).} \qquad (1)$$

For the optimal-growth closed-economy model (Ramsey-Cass-Koopmans), the reduced form convergence parameter $\beta$ depends on six structural parameters: the same four technological parameters as in the Solow-Swan model and two utility parameters of the representative generations of consumers. Utility and preferences of generations of the representative consumer include the discount rate over generations, $\rho$, and the relative fluctuation aversion parameter, $\theta$, (Barro and Sala-I-Martin 2004, 111):

$$\beta\big((\alpha, x, n, \delta), (\rho, \theta)\big)$$
$$= \tfrac{1}{2}\left\{\zeta^2 + 4(1-\alpha)\left(\tfrac{\rho+\delta+\theta x}{\theta}\right)\left[\left(\tfrac{\rho+\delta+\theta x}{\alpha}\right) - (n + x + \delta)\right]\right\} - \tfrac{\zeta}{2}, \qquad (2)$$
$$\zeta = \rho - n - (1-\theta)x > 0.$$

In the case when $(\rho + \delta + \theta x)/\theta = \alpha(n + x + \delta)$, the optimal-growth convergence parameter is identical to the constant savings-rate convergence parameter (Barro and Sala-I-Martin 2004, 109).

To identify the model a number of assumptions and additional estimations have been proposed: Ramsey (1928) argues against discounting future generations on an ethical principle of equality among generations which leads to the assumption of $\rho = 0$. The growth rate of the population $n$ can be easily computed. The growth rate of labour-augmenting technical change $x$ can be found by estimating a production function, although the measure of the stock of any type of capital (physical or human) depends on judgments on the depreciation of the variety of capital goods, $\delta$.

This leaves two structural parameters to be identified, namely $\alpha$, measuring the curvature of the production function, and $\theta$, measuring the curvature of the utility function. They indicate the degree of concavity or the extent to which the production function and the utility function differ from a linear function. The first one, $\alpha$, is the return to scale of any kind of capital that can be accumulated without instant depreciation. When the returns to scale tend to one, convergence is extremely slow. The second one, $\theta$, is the relative aversion of fluctuations of consumption (the inverse of the intertemporal elasticity of substitution). When the relative aversion to fluctuations tends to infinity, convergence is extremely slow.

The design of the convergence test does not allow to distinguish utility curvature $\theta$ from production curvature $\alpha$. There is observational equivalence for a given convergence parameter $\hat{\beta} = 2\%$: it may correspond to large returns to capital with a low relative fluctuation aversion and conversely. Because of under-identification with too many structural parameters, the convergence hypothesis test *cannot* test the exogenous savings rate (Solow growth model) *versus* endogenous savings (Ramsey, Cass, Koopmans optimal growth model).

The key achievement of *normative* optimal growth theory (utility characteristics matter) is forgotten in the *positive* convergence hypothesis empirical literature, which is nonetheless described in textbook to be the closest empirical test of neoclassical growth theory (Barro and Sala-I-Martin 2004, 466). The contribution of Ramsey-Cass-Koopmans-Malinvaud optimal growth theory is to state that utility matters due to preferences for smoothing consumption over time and generations. *Utility matters in normative theory but not in the positive theory testing the convergence hypothesis.*

## 4. ADAPTIVE EXPECTATIONS VAR VERSUS RATIONAL EXPECTATIONS DSGE

### *4.1. Two competing approaches*

In this section we will analyze the problem of observational equivalence and identification for DSGE models in more detail. To do this, we will compare the DSGE approach with adaptive expectations vector auto-regressive models. The econometric estimations of systems of dynamic equations of macroeconomic time-series, named vector auto-regressive (VAR) models, expanded among others into structural VAR (SVAR, Kilian and Lütkepohl 2017), cointegrated VAR (CVAR), panel VAR and global VAR (GVAR). These approaches, which are also dynamic, stochastic, general and equilibrium models, provide competing empirical macroeconomic explanations.

Most of the SVAR published papers refer to adaptive expectations 'simple' macroeconomic theories when explaining the statistical significance of impulses responses functions. SVAR is not a particular macroeconomic theory. It is a statistical method to test alternative specifications and restrictions of existing macroeconomic theories against each other. There remains a need for SVAR «*to bridge the gap between theory and data by developing structural models beyond the ones associated with data-induced restrictions*» (Spanos 2009, 11). Inventing these theories may be grounded on cost-benefit analysis (microeconomic foundations), on rule-based behaviour (agent-based models) or any other criteria different than only fitting CVAR.

Juselius and Franchi (2007), Spanos (2009) and Poudyal and Spanos (2016) contrast a 'theory first' or 'pre-eminence of theory' approach and a 'data first' approach. In a 'theory first' approach the facts *should* adjust to the theory such as in rational expectations New-Keynesian DSGE theory. In a 'data first' approach such as in structural or co-integrated vector auto-regressive models (SVAR or CVAR) the specification of the macroeconomic theory has to change due to econometric results found using data. Juselius and Franchi's (2007) replication of Ireland's (2004) real business cycle (RBC) model and Poudyal and Spanos' (2016) replication of Smets and Wouters (2007) show how a careful investigation of the properties of time-series lead to suggestions to change the specifications of the corresponding DSGE theory considerably.

Central banks' research departments and academia use both, DSGE models and VAR models. In the private banking sector, however, VAR and other econometric methods are employed for forecasting rather than New-Keynesian DSGE models. Colander (2009) asks why New-Keynesian DSGE theory is 'more successful' as a theory-first approach than data-first approach of co-integrated vector auto-regressive models (C-VAR) or structural VAR models. He suggests that New-Keynesian DSGE theory may not require 'judgment' and may be more fitted for natural selection in academic journals and the academic labour market and academic 'social replication'.

But empirical work and VAR also qualify for academic and central banks' research department careers, as much as New-Keynesian DSGE theory. A *p*-value below 5% leads to statistical significance for parameters or impulse response functions is easy to reach with large sample and multiple testing. It does not require judgment as a criterion for publishing a research paper (Wasserstein and Lazar 2016). Colander's (2009) argument of 'social replication' explanation for the success of DSGE models is not so obvious.

The success of the VAR approach, as well as empirical research pooling cross-country macroeconomic time-series, another data-first approach, is mostly explained by technical change. The increased computational capacities, the instantaneous and free access of economic data on the web (without the time-consuming issue of typing them), the ease of use of statistical software with online help, and open source free software, created a crowd of young applied econometricians. All graduate students in economics are able to obtain quickly statistical results which were primarily the production of near-genius Nobel-prize level researchers in the 30s-60s and then of engineer-like nerds using mainframe computers in the 70s-80s. Technological change also benefited DSGE models, especially with the Dynare software, but to a lesser extent. Overall, this decreased the market share of academic publications of small-size closed-form macroeconomic theoretical models which do not require simulations. This market share was very large before Kydland and Prescott's (1982) real business cycle simulations.

In addition to Colander's 'social replicator', a number of arguments can be found to explain why neither VAR nor DSGE managed to eradicate the other approach from macroeconomic research.

There is no conclusive empirical test of adaptive expectations against rational expectations. For every linear rational expectations DSGE model, there is an observationally equivalent adaptive expectations VAR model. *Data and econometrics cannot and will never be able to settle the controversy between adaptive and rational expectations.*

Some models, however, are better specified than others to match macroeconomic data. RBC models, since they include only structural parameters, are misspecified with respect to macroeconomic time-series persistence. The time series are often auto-correlated with two lags (AR(2)), with hump-shaped impulse response functions following shocks. DSGE modelers shifted to a hybrid model of rational expectations DSGE and (*ad hoc*) adaptive expectations VAR.

The adaptive expectations part of the DSGE model deals with the auto-correlation of an order at least equal to two of the observable macroeconomic time-series. Each observable time-series depends linearly on *ad hoc* (free) parameters for its lags (for example, habit persistence and inflation indexation) and for the auto-correlation parameter of its exogenous shock. It is assumed that these reduced-form parameters do not depend on policy-rule parameters. This assumption rules out Lucas' (1976) critique for the modelling of persistence in DSGE models. This adaptive-expectations part is observationally equivalent to an adaptive-expectations VAR including at least two lags without assuming the auto-correlation of the exogenous shocks. The forecasting performance of this hybrid DSGE is widely driven by its adaptive-expectations VAR part, sometimes up to 99%, when the observed macroeconomic time-series auto-correlations are close to unit roots. Coincidentally, at least half of the estimates of the auto-correlation parameters of shocks are usually close to unit root (between 0.95 and 0.998) in DSGE models.

The rational expectations part of the DSGE model mostly deals with the cross correlation among macroeconomic time series. The structural rational expectations part of the model has a marginal share of the forecasting ability of the model. To check this, one needs only to get back to the misspecified model setting all persistence free parameters to zero. In this part, reduced form parameters are highly non-linear functions of structural parameters. Hence, it is especially this part that contributes to the opacity of the weak identification issues.

With observational equivalence and the lack of identification or weak identification, a researcher does not receive a signal from a statistical software when there is a parameter identification problem while the parameters (using the *t*-test) may be significant. A weakly and/or not identified theory is likely to be successful, because the desired parameters are more easily statistically significant (any value may actually be estimated in some cases). The identification opacity allows the model to be 'not even wrong' with respect to its estimations of structural parameters. This leads to Gresham's law for preferring non-identified and weakly identified theories, a topic that will be discussed later in section 5.

The solutions of rational expectations DSGE models and their interpretations are way more complicated than the ones of adaptive expectations models. Opaque identification issues of structural parameters do not arise with adaptive-expectations VAR models. Ockham's razor argument is to seek the simplest model for observationally equivalent models with the simplest explanation. This is the reason why DSGE models, despite being grounded on the cost-benefit analysis of microeconomic theory, did not eradicate VAR models. The results of VAR models are relatively easy to interpret as opposed to DSGE models. Their identification issues are less opaque (Kilian and Lütkepohl 2017). Rational expectations and micro foundations are an unnecessary hypothesis for VAR modelers.

*4.2. Observational equivalence*

To show the observational equivalence of adaptive and rational expectations, consider the following class of hybrid linear macroeconomic DSGE models as stated in Koop *et al.* (2013):

$$\boldsymbol{A}_0(\theta)\boldsymbol{y}_t = \boldsymbol{A}_1(\theta)\boldsymbol{E}_t(\boldsymbol{y}_{t+1}) + \boldsymbol{A}_2\boldsymbol{y}_{t-1} + \boldsymbol{A}_3(\theta)\boldsymbol{x}_t + \boldsymbol{u}_t,$$

$$\boldsymbol{x}_t = \boldsymbol{\phi}_x \boldsymbol{x}_{t-1} + \boldsymbol{v}_t \text{ and } \boldsymbol{u}_t = \boldsymbol{\phi}_u \boldsymbol{u}_{t-1} + \boldsymbol{\varepsilon}_t. \quad (3)$$

The parameters of the model, $\theta$, are derived from the underlying theoretical model, e.g. maximization of utility and technology; they are sometimes called deep parameters. Endogenous variables are denoted $\boldsymbol{y}_t$, observable exogenous variables $\boldsymbol{x}_t$, non-observable exogenous variables $\boldsymbol{u}_t$, and independently and identically non-correlated shocks $\boldsymbol{v}_t$ and $\boldsymbol{\varepsilon}_t$. Endogenous aggregate variables depend on rational expectations $\boldsymbol{E}_t(\boldsymbol{y}_{t+1})$ for a proportion of of economic agents and on adaptive expectations $\boldsymbol{y}_{t-1}$ for the remaining proportion of economic agents. The reduced form parameters are given in matrices $\boldsymbol{A}_0(\theta)$, $\boldsymbol{A}_1(\theta)$, $\boldsymbol{A}_3(\theta)$, depending on the parameters $\theta$, and exogenous matrices $\boldsymbol{A}_2$, $\boldsymbol{\phi}_x$, and $\boldsymbol{\phi}_u$. An important restriction, common to most of the estimated DSGE models, is that the stock-flow equations of endogenous state variables (the stocks of wealth, capital, public and private debt) including the feedback rules of optimal control by economic agents are replaced by the auto-correlation equations of exogenous predetermined variables $\boldsymbol{x}_t$ and $\boldsymbol{u}_t$.

Assuming $\boldsymbol{A}_0(\theta)$ to be non-singular and using the method of undetermined coefficient, the solution of this model is (Binder and Pesaran 1997):

$$\boldsymbol{y}_t = \boldsymbol{C}(\theta)\boldsymbol{y}_{t-1} + \boldsymbol{G}_1(\theta, \boldsymbol{\phi}_x)\boldsymbol{x}_t + \boldsymbol{G}_2(\theta, \boldsymbol{\phi}_u)\boldsymbol{u}_t \quad (4)$$

where $\boldsymbol{C}(\theta)$ is a solution of the following quadratic equation:

$$A_1(\theta)C(\theta)^2 - A_0(\theta)C(\theta) + A_2 = 0. \tag{5}$$

Using the assumption that observable and unobservable exogenous variables follow a VAR(1), this solution is *observationally equivalent to a VAR(1) where expectations are adaptive*. Pesaran (1981, 376) shows that «in the absence of a priori restrictions on the processes generating the exogenous variables and the disturbances, the rational expectations and the general distributed lag models will be observationally equivalent; therefore, the auto-regressive and rational methods of expectations formation, cannot be distinguished from each other empirically.» For any rational expectations DSGE model, there is an observationally equivalent adaptive expectations VAR model. There is no conclusive test of adaptive *versus* rational expectations. To enlighten the point further the example of inflation persistence is analyzed in more detail

*4.3. Inflation persistence*

### 4.3.1. Adaptive expectations VAR models

Five different models with adaptive expectations are distinguished.

(1) Model $\mathcal{M}_{A1}$ is the Old-Keynesian AR(1) theory of inflation persistence which assumes backward-looking adaptive expectations so that inflation $\pi_t$ is an auto-regressive process of order 1 with an auto-correlation parameter $\lambda$ and a disturbance $\varepsilon_t$:

$$\mathcal{M}_{A1}: \pi_{t+1} = \lambda\pi_t + \varepsilon_{t+1} \text{ where } 0 < \lambda < 1 \text{ with } \pi_0 \text{ given, and } \varepsilon_t \, i.i.d.\, N(0,\sigma^2)$$

$$\Rightarrow E_t\pi_{t+1} = \lambda\pi_t.$$

The initial value of predetermined inflation $\pi_0$ is given. Inflation expectations are adaptive, they depend only on past values of inflation.

(2) Model $\mathcal{M}_{A2}$ is an auto-regressive ($\rho \neq 0$) disturbances or latent variable $u_t$ model. It is observationally equivalent to $\mathcal{M}_{A1}$.

$$\mathcal{M}_{A2}: \pi_{t+1} = u_{t+1} \text{ and}$$

$$u_{t+1} = \rho u_t + \varepsilon_{t+1}, \text{ where } 0 < \rho < 1$$

$$\text{with } \pi_0 \text{ given, and } \varepsilon_t \, i.i.d.\, N(0,\sigma^2)$$

$$\Rightarrow \pi_{t+1} = \rho\pi_t + \varepsilon_{t+1} \text{ and } E_t\pi_{t+1} = \rho\pi_t.$$

(3) Model $\mathcal{M}_{A3}$ is an AR(2) model of inflation with white-noise disturbances:

$$\mathcal{M}_{A3}: \pi_{t+1} = (\lambda + \rho)\pi_t - (\lambda\rho)\pi_{t-1} + \varepsilon_{t+1}.$$

(4) Model $\mathcal{M}_{A4}$ is an AR(2) model of disturbances. It is observationally equivalent to model $\mathcal{M}_{A3}$:

$$\mathcal{M}_{A4}: \pi_{t+1} = u_{t+1} \text{ and}$$

$$u_{t+1} = (\lambda + \rho)u_t - (\lambda\rho)u_t + \varepsilon_{t+1}.$$

(5) Model $\mathcal{M}_{A5}$ includes a lagged dependent variable along with an auto-regressive ($\rho \neq 0$) disturbances model $\varepsilon_t$.

$$\mathcal{M}_{A5}: \pi_{t+1} = \lambda\pi_t + u_{t+1} \text{ and}$$

$$u_{t+1} = \rho u_t + \varepsilon_{t+1}.$$

It is observationally equivalent to model $\mathcal{M}_{A3}$:

$$\pi_{t+1} - \lambda \pi_t = \rho(\pi_t - \lambda \pi_{t-1}) + \varepsilon_{t+1}$$

$$\pi_{t+1} = (\lambda + \rho)\pi_t - \lambda \rho \pi_{t-1} + \varepsilon_{t+1}.$$

Inverting $\lambda$ and $\rho$ is observational equivalent. Hence, if $(\rho \neq \lambda)$, there is an identification problem between the following two solutions for $\lambda$ and $\rho$:

$$\lambda = \frac{S + \sqrt{S^2 - 4P}}{2}, \rho = \frac{S - \sqrt{S^2 - 4P}}{2} \text{ or}$$

$$\lambda = \frac{S - \sqrt{S^2 - 4P}}{2}, \rho = \frac{S + \sqrt{S^2 - 4P}}{2}$$

with $S = \lambda + \rho$ and $P = \lambda \rho$.

Introducing both, a lagged dependent variable and an auto-regressive shock (as it is commonly done in DSGE models), leads to an identification issue of the auto-correlation of the shock. Furthermore, this model is observationally equivalent to an AR(2) model of the endogenous variable, with zero auto-correlation of the shock. Hence, any DSGE model assuming auto-regressive shocks in each equation (e.g. Smets and Wouters 2007) is observationally equivalent to another DSGE model assuming one more lag of each dependent variable *without an auto-regressive part* in the shocks of each equation.

### 4.3.2. Rational expectations DSGE models of persistence

Key element of the New-Keynesian theory of inflation persistence is the New-Keynesian Phillips curve (Galí 2015). Inflation is assumed to be forward-looking, without an initial condition for inflation. Lubik and Schorfheide (2006) and An and Schorfheide (2007) consider two observationally equivalent rational expectations DSGE models of inflation. For both models, inflation $\pi_t$ is the observed endogenous forward-looking variable with potentially exploding dynamics where the growth factor $1/\beta > 1$ is driven by inflation expectations $E_t \pi_{t+1}$. Both models include independently and identically distributed shocks $\varepsilon_t$. Without loss of generality, we will assume in the following analysis that the output gap does not deviate from zero and that there is no Taylor rule.

(1) Model $\mathcal{M}_{R1}$ includes a non-observable backward-looking auto-regressive (forcing variable) cost-push shock $u_t$ with an auto-correlation parameter $0 < \rho < 1$, with a given initial value $u_0$.

$$\mathcal{M}_{R1}: \pi_t = \beta E_t \pi_{t+1} + u_t, 0 < \beta < 1,$$

$$u_t = \rho u_{t-1} + \varepsilon_t, 0 < \rho < 1.$$

Initial inflation is not given in model $\mathcal{M}_{R1}$.

(2) In the hybrid New-Keynesian Phillips curve model $\mathcal{M}_{R2}$ current inflation depends not only on expectations about future inflation, but also on lagged inflation with a parameter $b$, whose boundaries are to be determined. Because it is assumed that a proportion of agents are backward-looking, initial inflation $\pi_0$ is given.

$$\mathcal{M}_{R2}: \pi_t = \beta E_t \pi_{t+1} + b \pi_{t-1} + \varepsilon_t,$$

$$0 < \beta < 1, 0 < b < b_{max}(\beta) < 1.$$

(3) Beyer and Farmer (2007) compare models $\mathcal{M}_{R1}$ and $\mathcal{M}_{R2}$ to model $\mathcal{M}_{R3}$:

$$\mathcal{M}_{R3}: \pi_t = aE_t\pi_{t+1} \text{ with } |a| > 1.$$

They demonstrate that the solution of multiple equilibria of this model depends on an i.i.d. sunspot shock $w_t$, thus exhibiting indeterminacy:

$$\mathcal{M}_{R3}: \pi_t = \frac{1}{a}\pi_{t-1} + w_t.$$

The indeterminacy solution of $\mathcal{M}_{R3}$ is observationally equivalent to the determinacy solution of $\mathcal{M}_{R2}$. Furthermore, Lubik and Schorfheide (2006) and An and Schorfheide (2007) demonstrate that the determinate solution of model $\mathcal{M}_{R2}$ is observationally equivalent to the one of model $\mathcal{M}_{R1}$. We will demonstrate in this section that the two determinate solutions of rational expectations models $\mathcal{M}_{R1}$, $\mathcal{M}_{R2}$ and the indeterminate solutions of model $\mathcal{M}_{R3}$ are observationally equivalent to the adaptive expectations model $\mathcal{M}_{A1}$. This last result is an example of Pesaran's (1981) observational equivalence result between rational and adaptive expectations.

(A) Observational equivalence between rational expectations model with latent auto-regressive variable $\mathcal{M}_{R1}$ and adaptive expectations model $\mathcal{M}_{A1}$: When transforming the model $\mathcal{M}_{R1}$ into its matrix form, the eigenvalues can be found on the diagonal of the transition matrix, one of them is unstable, $1/\beta$, the other one stable, $\rho$:

$$\begin{pmatrix} E_t\pi_{t+1} \\ u_{t+1} \end{pmatrix} = \begin{pmatrix} \frac{1}{\beta} & -\frac{1}{\beta} \\ 0 & \rho \end{pmatrix} \begin{pmatrix} \pi_t \\ u_t \end{pmatrix} + \begin{pmatrix} 0 \\ 1 \end{pmatrix}\varepsilon_t.$$

According to Blanchard and Kahn's (1980) unique solution for rational expectations models, inflation has to be exactly collinear to the non-observable shock with the parameter $N$ corresponding to the slope of the eigenvector of the stable eigenvalue:

$$E_t\pi_{t+1} = \rho\pi_t = \frac{1}{\beta}\pi_t - \frac{1}{\beta}u_t \Rightarrow \pi_t = \frac{1}{\rho - \frac{1}{\beta}}\frac{-1}{\beta}u_t = Nu_t, \text{with } N = \frac{1}{1 - \beta\rho} > 1.$$

Since the shock $u_t$ is not observable it is not possible to run the regression $\pi_t = Nu_t$, which is predicted to be exact, with a coefficient of determination equal to 1 and with a negative slope ($N < 0$). What is observable, however, is the auto-correlation of inflation.

$$\pi_{t+1} = Nu_{t+1} = N\rho u_t + N\varepsilon_{t+1} = \frac{N\rho\pi_t}{N} + N\varepsilon_{t+1} = \rho\pi_t + \frac{1}{1-\beta\rho}\varepsilon_{t+1} \Rightarrow E_t\pi_{t+1} = \rho\pi_t.$$

But in the above equation, the parameter $1/\beta$, related to expectations, only appears in the variance of the disturbances, which does not allow to identify $1/\beta$. The econometrician cannot distinguish between the adaptive expectations Old-Keynesian residuals $\varepsilon_{t+1}$ and the rational expectations New-Keynesian residuals multiplied by $N$: $N\varepsilon_{t+1}$. The model $\mathcal{M}_{R1}$ is observationally equivalent to the adaptive expectations VAR model $\mathcal{M}_{A1}$. Table 1 summarizes which variable is observed and which auto-correlation parameter can be estimated in model $\mathcal{M}_{R1}$.

**Table 1:** The Chiasm of the New-Keynesian Synthesis in model $\mathcal{M}_{R1}$

|  | Observed | Not observed |
|---|---|---|
| Variable | Inflation | Cost-push forcing variable |
| Auto-regressive parameter | Cost-push shock: $0 < \rho < 1$ | Inflation: $\frac{1}{\beta} > 1$ |

(B) Observational equivalence between the hybrid rational expectations and adaptive expectations model $\mathcal{M}_{R2}$ and the adaptive expectations model $\mathcal{M}_{A1}$.

To find the unique solution of the **hybrid model** $\mathcal{M}_{R2}$, we have to solve:

$$0 = P(\mu) = \beta\mu^2 - \mu + \beta \text{ assuming } 0 < \Delta = 1 - 4b\beta \Rightarrow b\beta < \frac{1}{4}$$

$$\mu_1(\beta, b) = \frac{1}{2\beta}(1 - \sqrt{1 - 4b\beta}), \quad \mu_2(\beta, b) = \frac{1}{2\beta}(1 + \sqrt{1 - 4b\beta}).$$

There is again one stable and one unstable eigenvalue. $P(1) = \beta - 1 + b < 0$, $(b < 1 - \beta)$ implies that $\mu_1 < 1 < \mu_2$. $P(-1) = \beta + 1 + b > 0$, $(b > -1 - \beta$, always satisfied if $b > 0$) implies that $-1 < \mu_1 < \mu_2$. For $b > 0$, $\beta > 0, \beta + b < 1$, (then $b\beta \leq \frac{1}{4}$ is satisfied), there is a unique rational expectations solution. The unit root for inflation ($\mu_1 = 1$) can be reached in the limit case $\beta + b = 1$ and $\beta < 0.5$ (Dees *et al.* 2009).

$$\pi_t = \mu(\beta, b)\pi_{t-1} + \frac{1}{1-\beta\mu(\beta,b)}\varepsilon_t. \tag{6}$$

For values of parameters $\rho = \mu(\beta, b)$, the auto-correlation of inflation and the variances of the random disturbances are identical for models $\mathcal{M}_{R2}$ and $\mathcal{M}_{R1}$ (Lubik and Schorfheide 2006). They are also observationally equivalent to an adaptive-expectations model $\mathcal{M}_{A1}$ with the same auto-correlation and the same variance of the shocks (An and Shorfheide 2007, Pesaran 1981). Lubik and Schorheide (2006) and An and Schorfheide (2007) state that both parameters $(\beta, b)$ are not identified, as only the reduced form parameter $\mu(\beta, b)$ can be estimated. One identifying restriction is required. Else, one can identify $(\beta, b)$ in another specification which includes at least one additional *observable* regressor which follows *an auto-regressive process at least of order 2*, such as the output gap (Dees *et al.* 2009) which for US data is weakly identified (Mavroeidis *et al.* 2011).

### 4.3.3. Criteria for selecting between observationally equivalent models

We suggest four criteria for selecting between the adaptive expectations model $\mathcal{M}_{A1}$ and the three observationally equivalent rational expectations DSGE models $\mathcal{M}_{R1}$, $\mathcal{M}_{R2}$, and $\mathcal{M}_{R3}$.

**Criterion 1**: Micro-foundations of rational behaviour and Lucas critique: All models are *ad hoc* exogenous models of persistence with free parameters for the auto-correlation $(\lambda, \rho, \mu)$. This property is valid independently of the number of equations of the hybrid DSGE model (exogenous auto-correlation matrices $\boldsymbol{A}_2$ and $\boldsymbol{\phi}_u$). They are unrelated to the micro-foundations of rational decisions of private sectors agent. They are assumed to be unrelated to a potential feedback effect of the policy rule of policy-makers. They may all face the Lucas critique.

**Criterion 2:** Parsimony (fewer parameters: Ockham's razor): This criterion is valid for normative and positive theory. Adaptive expectations with only one parameter is better than rational expectations with two parameters.

**Criterion 3:** Identified parameter instead of non-identified parameters for the positive theory: This criterion is valid for a positive theory only. Adaptive expectations have an identified parameter (the number of structural parameters is equal to the number of reduced form parameters), whereas rational expectations have non-identified parameters, requiring at least one identifying restriction.

**Criterion 4:** A more *simple* explanation and interpretation for adaptive expectations with respect to a more *complicated* explanation for rational expectations models: The simplicity of explanations and interpretations as part of Ockham's razor argument.

In model $\mathcal{M}_{R2}, b = \mu - \beta\mu^2 < \mu$, with an extreme gap $\mu - b$ is obtained for around $\beta = b = 0.5$ leading to $\mu = 1$. The auto-correlation $b$ of the the non-observable proportion of adaptive expectations agents has a non-linear effect on the observed auto-correlation of inflation, while being always smaller than this observed auto-correlation (figure 1). These effects of the hybrid models are *more complicated* than the observationally equivalent model with *all* agents following adaptive expectations where $\mu = b$.

**Figure 1**: Observed inflation auto-correlation as a function of its auto-correlation $b$ in the hybrid model, for its dependence on rational expectations agents: $\beta = 0.6$ (red), 0.5 (purple), 0.9 (dark green) and 0 (light green, adaptive expectations).

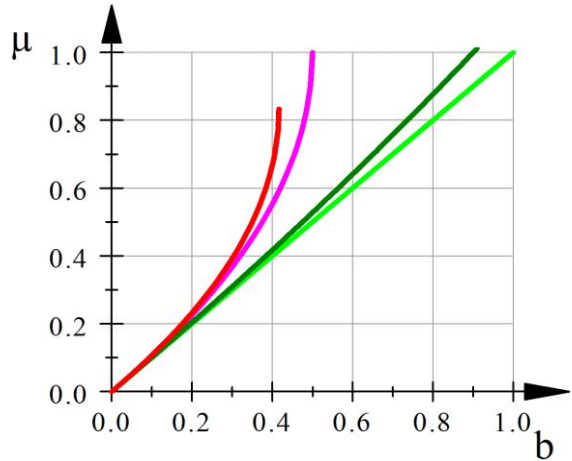

Model $\mathcal{M}_{R3}$ with its indeterminate solutions (sunspots) is observationally equivalent to the other models with determinacy (Beyer and Farmer 2007). Taking into account sunspots solutions of indeterminacy complicates even more the interpretations of rational expectations hypothesis.

In model $\mathcal{M}_{R1}$, the observed initial value of inflation is anchored by private sector agents on a cost-push variable $\pi_0 = Nu_0$. *Private sector's agents exactly know the current and future values of this auto-regressive cost-push shock $u_t$ and, at the same time, they cannot observe them*. Anywhere else than in New-Keynesian macroeconomic theory, this would be an anchor that fails, leading to many sunspots, because it is not observed and therefore, imperfectly known. In this theory, it is the reverse: a non-observable shock is unique and avoids the occurrence of multiple equilibria for initial conditions (sunspots). The non-observable anchor $u_0$ of rational expectations New-Keynesian model $\mathcal{M}_{R1}$ has a similar status for the adaptive expectations model $\mathcal{M}_{A2}$ as the existence of non-observable God in Laplace's celestial mechanics. Adding one, two or an infinity of auto-regressive non-observable latent variables leading to observationally equivalent models is always possible. It may be *useful for normative theory* and *useless for positive theory*. This is an *unlikely and complicated* explanation which is not required in two other observationally equivalent models. Initial inflation $\pi_0$ is given in the adaptive expectations Keynesian model $\mathcal{M}_{A1}$ and for the hybrid model $\mathcal{M}_{R2}$ including a proportion of agents with adaptive expectations.

### 4.4. Parameter identification issues

Observationally equivalent models do not necessarily imply a lack of identification of one of their parameters. As seen in the former examples, two models may be observationally equivalent while both models have all their parameters identified, while one or more parameters are non-identified in one model, whereas all are identified in the other model, or while having one or more parameters non-identified in both models.

Parameters relating current variables to their expectations are not identified in a number of models by Beyer and Farmer (2004) as has been pointed out by Cochrane (2011). This will be clarified in the following two examples starting again with the hybrid linear model analyzed before:

$$\boldsymbol{A}_0(\theta)\boldsymbol{y}_t = \boldsymbol{A}_1(\theta)\boldsymbol{E}_t(\boldsymbol{y}_{t+1}) + \boldsymbol{A}_2\boldsymbol{y}_{t-1} + \boldsymbol{A}_3(\theta)\boldsymbol{x}_t + \boldsymbol{u}_t,$$

$$\boldsymbol{x}_t = \boldsymbol{\phi}_x\boldsymbol{x}_{t-1} + \boldsymbol{v}_t \text{ and } \boldsymbol{u}_t = \boldsymbol{\phi}_u\boldsymbol{u}_{t-1} + \boldsymbol{\varepsilon}_t.$$

**Example 1:** $\boldsymbol{A}_2 = 0$, $\boldsymbol{\phi}_x = 0$, $\boldsymbol{\phi}_u = 0$.

With neither lags, nor auto-regressive exogenous regressors, the model is a degenerate rational expectations model without predetermined variable. Its unique solution is observationally equivalent to a random error model without dynamics and expectations:

$$\boldsymbol{A}_0(\theta)\boldsymbol{y}_t = \boldsymbol{A}_1(\theta)\boldsymbol{E}_t(\boldsymbol{y}_{t+1}) + \boldsymbol{\varepsilon}_t \Rightarrow \boldsymbol{A}_0(\theta)\boldsymbol{y}_t = \boldsymbol{\varepsilon}_t \Rightarrow \boldsymbol{y}_t = \boldsymbol{A}_0(\theta)^{-1}\boldsymbol{\varepsilon}_t \qquad (7)$$

The solution does not depend on $\boldsymbol{A}_1(\theta)$. If some of the parameters denoted $\theta_1$ appear only in $\boldsymbol{A}_1(\theta_1, \theta_0)$ and not in $\boldsymbol{A}_0(\theta_0)$, they are not identified. They are related to the out-of-equilibrium unstable eigenvalues.

**Example 2:** $\boldsymbol{A}_2 = 0$, $\boldsymbol{\phi}_x = 0$, $\boldsymbol{\phi}_u = \rho \boldsymbol{I}_n$.

Without lags, but with auto-regressive unobservable exogenous regressors, the model includes $n$ forward-looking variables related to observable variables $y_{it}, i \in \{1, ..., n\}$ and $n$ corresponding non-observable predetermined exogenous auto-regressive forcing variables $\boldsymbol{u}_t$. The matrix $\boldsymbol{\phi}_u = diag(\rho_i)$ is a diagonal matrix of exogenous auto-correlation coefficients $\rho_i, |\rho_i| < 1$. Hence, all the eigenvalues of the auto-correlation matrix $\boldsymbol{\phi}_u$ are stable. Blanchard and Kahn's (1980) determinacy condition implies that the remaining eigenvalues of the matrix $\boldsymbol{A}_1(\theta)^{-1}\boldsymbol{A}_0(\theta)$ related to the endogenous part of the model should all be unstable. Forward-looking variable are linearly anchored on non-observable forcing variables, with a unique matrix $\boldsymbol{N}(\theta)$:

$$\boldsymbol{y}_t = \boldsymbol{N}(\theta)\boldsymbol{u}_t = \boldsymbol{N}(\theta)\boldsymbol{\phi}_u\boldsymbol{u}_{t-1} + \boldsymbol{N}(\theta)\boldsymbol{\varepsilon}_t = \boldsymbol{N}(\theta)\boldsymbol{\phi}_u\boldsymbol{N}(\theta)^{-1}\boldsymbol{y}_{t-1} + \boldsymbol{N}(\theta)\boldsymbol{\varepsilon}_t$$

$$\boldsymbol{y}_t = \rho\boldsymbol{y}_{t-1} + \boldsymbol{N}(\theta)\boldsymbol{\varepsilon}_t \text{ if } \boldsymbol{\phi}_u = \rho\boldsymbol{I}_n.$$

When all the auto-correlation coefficients are identical, the model is observationally equivalent to an *ad hoc* adaptive expectations model only driven by the exogenous auto-regressive parameter $\rho$, upon which no economic agent has any influence, despite their rational optimizing behaviour.

When the exogenous auto-correlation parameters are different ($\rho_i \neq \rho_j$) as in Smets and Wouters (2007), the model is related to economic behaviour only through the matrix $\boldsymbol{N}(\theta)$ in the hope that identification of all parameters $\theta$ is feasible, which is not always granted. The economic interpretation of the effects $N(\theta)\boldsymbol{\phi}_u N(\theta)^{-1}$ of structural parameters $\theta$ on observable variables $\boldsymbol{x}_t$ is opaque.

## 5. GRESHAM'S LAW FOR THE THEORY OF POSITIVE MACROECONOMICS

### *5.1. Against the method of positive economics*

«*What do I think of AM [Against Method] today? Well, scientist have always acted as a loose and rather opportunistic way when doing research, though they have often spoken differently when pontificating about it. By now, this has become a commonplace among historians of science.*» (Feyerabend 1995, 151)

When bringing the normative, theory-first, Walrasian New-Keynesian DSGE model to the data, an outcome of this theoretical rigor where '*very few things go*' is an empirical strategy which avoids a kind of Popperian falsification of theory-first models, while pretending to do this falsification. It uses observational equivalence and the opacity of parameter identification. This empirical and inductive poor method is observationally equivalent to ('as if') a kind of '*many things go*' lack of scientific method on its empirical side. This 'many things go' of the empirical side of positive theory is the flip side of testing the strict constraints of 'very few things go' of normative theory.

Observational equivalence, under-identification and weak identification issues defend the academic freedom to try, to assume and to defend any unlikely or unnecessary hypothesis which are considered to be interesting for other criteria than the ability to be falsified in positive economics.

But any unlikely and unnecessary hypothesis does not come from anywhere in mainstream macroeconomics. The methodological anarchism of 'many things go' for positive economics, undercover of a seemingly Popperian falsification, is reserved to the happy few in the hierarchy of academic power. They have the academic freedom to choose the 'very few things go' of the normative theory and the 'many things go' when bringing it to the data as a positive theory. Mainstream macroeconomics is a highly top-down hierarchical field of research where the social norm of what is accepted for publication in best academic journals at a given point in time is decided in top-10 macroeconomics departments, in the NBER summer institute macroeconomics meetings, and in Fed and ECB research departments. For now, ten years since the great financial crisis of 2007, there are recurrent claims against a lack of pluralism of macroeconomic theory, where New-Keynesian DSGE models are the only game in town. Empirical methodological anarchism for the few is compatible with the lack of theoretical methodological pluralism for the many.

Treating openly identification issues leads to modest and disappointing results. It may lead to the empirical rejection of New-Keynesian DSGE theory. It is rewarding for the scientific careers of theory-first macroeconomists to escape a detailed treatment of identification issues. While pretending empirical evidence and Popperian falsification, this strategy maintains theory-first Walrasian tradition alive, doing business as usual, with the help of step 1:

(1) Claim the empirical evidence of some of the propositions derived from the axioms. The evidence is grounded on a biased interpretation of observationally equivalent models facing exact and weak identification issues.

(2) Deductive theory instead of inductive.

(3) Normative as if positive.

(4) Coordinate on the axioms of the deductive theory colluding with a network of allies and involve a network of stakeholders, such as central bank research departments.

(5) Defend the axioms of the deductive theory using arguments of authority and hierarchy, ideology, political policy rhetoric, in order to force dissenters out of the field of mainstream macroeconomics (Romer 2016).

*5.2. Opacity and 'mathiness'*

This normative-positive empirical approach may go hand in hand with Romer's (2015) 'mathiness' blurring the interpretation and the understanding of economic theory. We may define mathiness as an ambiguous literary discourse, analogies and interpretation of formal mathematical models, with a narrative which describes something else or the opposite of the working of the formal mathematical model, to get it published, to sell the paper in broadening its contribution and its empirical evidence, to convince, and to find new allies.

'Mathiness' in New-Keynesian DSGE modelling is related to technological progress: the development of the Dynare software in the 2000s. While easing access to DSGE simulations and Bayesian econometrics to new PhD students, many of this DSGE-born generation of macroeconomists did not acquire the knowledge and the modeling skills of small scale consistent macroeconomic models and the understanding of their mathematical solutions. This widened the gap between their literary fairy-tale narrative on the mechanisms underlying the impulse response functions of their DSGE models and their effective, obscure, and messy mechanisms. These mechanisms are related to several distortions with respect to perfect competition equilibrium, while omitting from time to time stock-flow accounting equations, assuming that variables such as the capital stock are not observable. «*It is often extremely hard to understand what a particular distortion does on its own and then how it interacts with other distortions in the model.* » (Blanchard 2016, 3).

*5.3. New-Keynesian DSGE withers*

Will the fate of the New-Keynesian DSGE theory in twenty years be different from that of the theories of the 80s? Although Bayesian estimations still deliver plausible parameters in many published DSGE papers as of today, the core five equations New-Keynesian DSGE model are at odds with macroeconomic facts of the last ten years describing a new accumulation regime for OECD countries following the great financial crisis of 2007.

First, the *key* transmission equation of monetary policy is the New-Keynesian Phillips curve. It plays a similar role as the quantitative theory of money in monetarism. In the 2010s, some researchers state that the New-Keynesian Phillips curve (as well as the old Keynesian Phillips curve) abandoned them (Mavroeidis *et al.* 2014, Borio 2017).The instability of the velocity of money for the monetary policy transmission mechanism was the argument to abandon monetarism in the 1990s. Mavroeidis *et al.* (2014) surveys the instability of the estimates of the slope of the new-Keynesian Phillips curve, which is the key parameter of the monetary transmission mechanism of the New-Keynesian DSGE theory. Borio (2017) emphasizes the zero correlation between inflation and output in the last ten years.

Secondly, the intertemporal substitution effect of the interest rate on future output is smaller than expected in the US and not different from zero in most of other countries in the world (Havranek *et al.* 2015). The habit persistence parameters are also found to be much smaller, and close to zero with microeconomic estimates (Havranek *et al*. 2017).

Thirdly, there is no evidence of the positive-feedback mechanism of Taylor rule within the New-Keynesian model because of its lack of identification (Cochrane 2011).

Fourthly, the assumption of zero net supply of public and private debt in Galí's (2015) book and the omission of the stock-flow debts and capital equations in DSGE econometrics does not fit with the increase of debt which occurred in the last thirty years (Borio 2017), altogether with the fall of long run interest rates.

Fifth, the auto-correlation of shocks can be replaced by adding lags of order two or more of observable variables, such as inflation, output gap and federal funds rate.

The new data from a new macroeconomic accumulation regime since 2007 requires new models and new policy advice. Macro-prudential DSGE models have not been convincing. New technical tools related to big data and enhanced computational power will appear. A new generation of researchers have less stakes in DSGE models. New and unexpected dynamic stochastic macroeconomic models, with or without equilibrium, will emerge. At the beginning, they will be rhetorically presented as a new consensus, upgrading New Keynesian DSGE theory, in order to find allies during the transition. But *they will not keep the five core New-Keynesian DSGE equations*. Thousands of New-Keynesian DSGE published papers will belong to the history of dead economic theories. There is no guarantee, however, that these new macroeconomic theories will take a better care of identification issues instead of seeking observational equivalence and identification opacity in order to be successful.

## 6. CONCLUSION

Our point is not to promote data-first against theory-first, but an honest balance between theory and data, as proposed by Spanos (2009). This leads to *modest* but *robust* empirical results, once identification and weak identification issues are taken into account. A theory which has not too many parameters has more chances to last than a theory based on specific micro-foundations involving several non-testable structural parameters and unexpected sign prediction, such as the New-Keynesian Phillips curve. Modest and robust results are not demanded by top academic journals. They are not rewarding for researchers.

As a consequence, testing between macroeconomic theories in order to settle controversies has made very little robust progress. For this reason, New-Keynesian DSGE theory is likely to face creative destruction as it happened to theories of the 80s.

A first cost of the New-Keynesian DSGE theory is related to a bias in the allocation of talent. A 'forced consensus' emphasizes that there is no alternative on the job market for current PhD candidates in macroeconomic theory than writing New-Keynesian DSGE models. Skeptical PhD candidates with alternative ideas shift to microeconomics or econometrics.

A second cost of the New-Keynesian DSGE theory is that, by its recurrent attempts to force consensus while avoiding to consider identification issues in empirical tests, it maintains macroeconomic theory as a science in the making, fostering endless controversies. Macroeconomic theory does not accumulate at least a few modest and robust results into Latour's 'ready-made science', in order to complement, if possible, the resilient undergraduate IS-LM model.

With the identification problem and observational equivalence, we recommend to be rather less dogmatic than more dogmatic about economic theories since, as J.M. Keynes put it in a letter to Roy Harrod: «Economics is a science of thinking in terms of models joined to the art of choosing models which are relevant to the contemporary world.» (Keynes 1938).

APPENDIX

**Table 2**: Citations per year (since and including publication year) in Google Scholar, 15th December 2017

| Reference | Year | 1970s | 1980s | 1990s | 2000s | 2010s |
|---|---|---|---|---|---|---|
| Friedman | 1960 | 13 | 21 | 41 | 66 | 101 |
| Friedman/Schwartz | 1963 | 33 | 63 | 127 | 246 | 448 |
| Barro/Grossman | 1971 | 25 | 44 | 22 | 23 | 34 |
| Dornbusch | 1976 | 23 | 94 | 129 | 204 | 182 |
| Malinvaud | 1977 | 25 | 75 | 39 | 29 | 37 |
| Kydland/Prescott | 1982 | | 35 | 114 | 235 | 359 |
| Clarida/Galí/Gertler | 1999 | | | | 294 | 383 |
| Woodford | 2003 | | | | 333 | 490 |
| Smets/Wouters | 2003 | | | | 168 | 335 |
| Smets/Wouters | 2007 | | | | 156 | 412 |

The Google scholar database includes many citations errors with citations selected in a list of working papers and journals which changes over time. Many working papers series of the eighties have not been numerated and are not included in the database. Online working paper series expanded to maturity in the decade 2000s. The current population of citing researchers in economics measured by registered researchers in Repec database is around 50.000. With respect to the eighties, this citing population increased markedly following the increase of the number of university students around the world and adding new researchers in economics of former communist countries such as China and Russia as well as researchers from emerging economies. Hence, 100 citations per year in the 1980s may correspond to 400 citations per year in the 2010s in Google scholar database for a similar citation impact in the macroeconomic research community at the time.

In the 1980s, citations are of comparable magnitude for monetarism (Friedman 1960, Friedman and Schwartz 1963), fixed-price disequilibrium (Barro Grossman 1971, Malinvaud 1977), *ad hoc* rational expectations models (Dornbush 1976) and real business cycles (Kydland and Prescott 1982).

In the 2010s, citations of new-Keynesian DSGE articles or books (Clarida, Galí, Gertler 1999, Woodford 2003, Smets and Wouters 2003, 2007) citations are at around 400 per year. This is also the case for Kydland and Prescott (1982), although genuine RBC models are rarely simulated nowadays. Interestingly, Friedman and Schwartz (1963) is still heavily cited, although money demand estimations run out of fashion in working papers of central banks research departments. This book is also an account of the great depression of the 1930s, with has a renewed interest after the great financial crisis following 2007. These citations are not necessarily meaningful. They should not be taken as precise measures of the academic impact of research.